\documentclass[a4paper]{jpconf}
\usepackage{graphicx}
\usepackage[frozencache]{minted}

\begin{document}
\title{The Awkward World of Python and C++}

\author{Manasvi Goyal$^1$, Ianna Osborne$^2$, Jim Pivarski$^2$}

\address{$^1$ Delhi Technological University, Delhi, India}
\address{$^2$ Princeton University, Princeton, NJ 08544, USA}

\ead {mg.manasvi@gmail.com}

\begin{abstract}
There are undeniable benefits of binding Python and C++ to take advantage of the best features of both languages. This is especially relevant to the HEP and other scientific communities that have invested heavily in the C++ frameworks and are rapidly moving their data analyses to Python. Version 2 of Awkward Array, a Scikit-HEP Python library, introduces a set of header-only C++ libraries that do not depend on any application binary interface. Users can directly include these libraries in their compilation instead of linking against platform-specific libraries. This new development makes the integration of Awkward Arrays into other projects easier and more portable, as the implementation is easily separable from the rest of the Awkward Array codebase. The code is minimal; it does not include all of the code needed to use Awkward Arrays in Python, nor does it include references to Python or pybind11. The C++ users can use it to make arrays and then copy them to Python without any specialized data types - only raw buffers, strings, and integers. This C++ code also simplifies the process of just-in-time (JIT) compilation in ROOT. This implementation approach solves some of the drawbacks, like packaging projects where native dependencies can be challenging. In this paper, we demonstrate the technique to integrate C++ and Python using a header-only approach. We also describe the implementation of a new LayoutBuilder and a GrowableBuffer. Furthermore, examples of wrapping the C++ data into Awkward Arrays and exposing Awkward Arrays to C++ without copying them are discussed.

\end{abstract}

\section{Introduction}

\begin{figure}[htp]
    \centering
    \includegraphics[width=16cm]{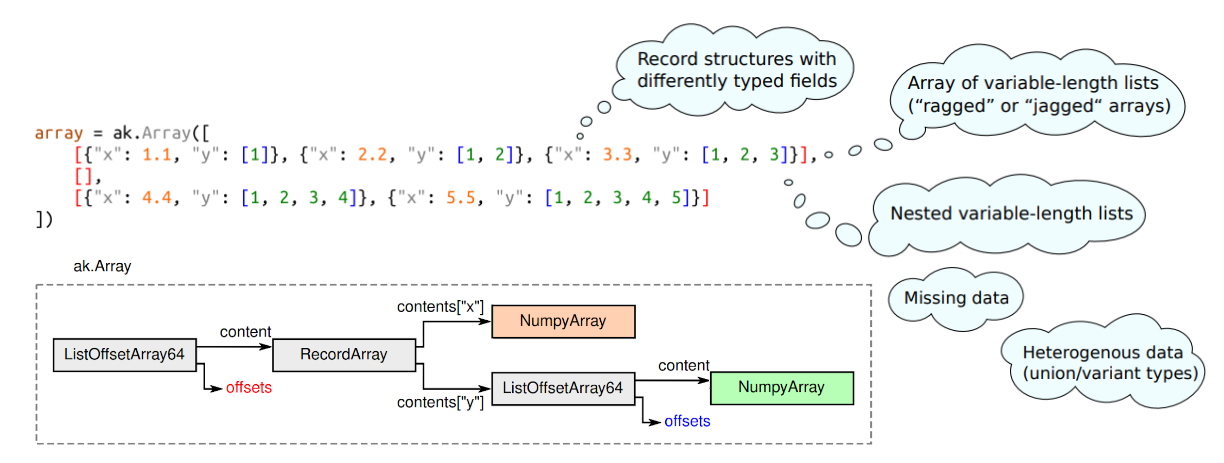}    \caption{Structure of an Awkward Array with nested variable-length lists and records, color-coded with an array example.}
    \label{fig:AwkwardArray_Structure}
\end{figure}

Awkward Array~\cite{awkward-ref} is an important tool for physics analysis in Python for High Energy Physics (HEP) community. It is part of the \verb"Scikit-HEP"~\cite{sci-kit-hep-ref} ecosystem. Nested, variable-length lists ("ragged" or "jagged" arrays), records with differently typed fields, missing data, and other heterogeneous data (union/variant types) can be defined as a set of primitives using NumPy-like~\cite{numpy-ref} phrases in Python~\cite{columnar-ref}. In Awkward arrays, a single user-facing \verb"ak.Array" consists of one small tree with large contiguous data buffers attached to each node~\cite{awkward-paper-ref}, as shown in Figure~\ref{fig:AwkwardArray_Structure}. Compiled operations are performed on these data buffers, not on the objects they represent.

In this work, we present new tools for creating Awkward Arrays in C++. Previously, the main codebase was written in C++~\cite{awkward-scipy2020-ref} with the idea that the downstream code would link to \verb"libawkward.so", but that route is full of hidden issues~\cite{awkward-lessonslearnt-ref}. The method of a small, header-only library that only fills array buffers for downstream code to pass from C++ to Python using only \verb"ctypes" has considerably more promise. 

\section{Python-C++ Integration}

Nowadays, more front-end users use Python~\cite{data-science-tools-ref}, but large-scale processing still needs to have high performance of C++~\cite{awkward-lessonslearnt-ref}. That is why we combine Python and C++ to take advantage of the best features of both languages so that we can have a Python user interface and, at the same time, take advantage of the performance and memory management of C++. HEP and other scientific communities have extensively invested in the C++ frameworks and are swiftly migrating their data analyses to Python. These communities are particularly interested in bridging the gap between the two languages~\cite{data-science-tools-ref}. This raises an important question: \textit{'How to do Python-C++ integration the right way?'}, which is addressed in the following sections.

\section{The 'Header-Only' Approach}

A set of header-only C++ libraries has been introduced to address the issues in the Python-C++ integration of Awkward Arrays~\cite{awkward-lessonslearnt-ref}. These templated C++ libraries are not dependent on any application binary interface. They can be directly included in a project's compilation without the need to link against platform-specific libraries. This 'header-only' approach not only simplifies the production of Awkward Arrays in a project but also enhances the portability of the Awkward Arrays. The code is minimal and does not constitute all of the code required to use Awkward Arrays in Python. It contains no references to Python or Python bindings. The header files can be used by C++ users to create Awkward Arrays, which can then be copied into Python without any specialized data types - only raw buffers, strings, and integers. This approach addresses the issue of packaging projects with native dependencies.

\section{LayoutBuilder}

\begin{figure}[htp]
    \centering
    \includegraphics[width=16cm]{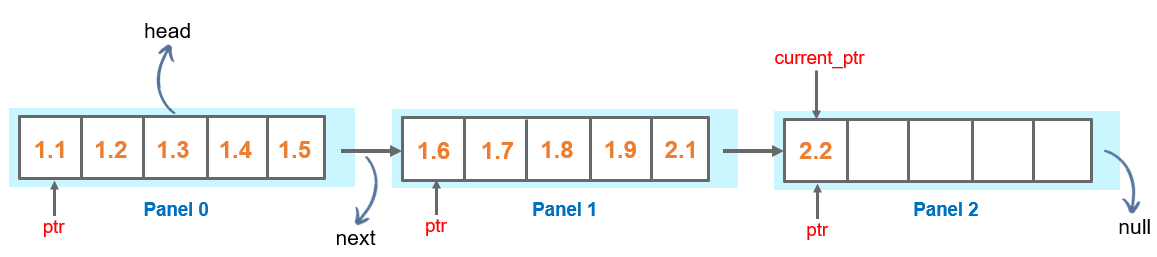}
    \caption{Awkward Array GrowableBuffer implemented as a linked list with multiple panels, each of size = 5, that are allocated as needed, i.e., when the GrowableBuffer runs out of space.}
    \label{fig:AwkwardArray_GrowableBuffer_ll}
\end{figure}

A 'layout' consists of composable elements that determine how an array is structured. It can only build a specific view determined by the layout Form. \verb"LayoutBuilder"~\cite{layoutbuilder-userguide-ref} is a set of compile-time, templated static C++ classes implemented entirely in a header-only library. It uses a header-only \verb"GrowableBuffer" (Figure~\ref{fig:AwkwardArray_GrowableBuffer_ll}), which is implemented as a linked list with smart pointers. \verb"awkward::LayoutBuilder" specializes an Awkward data structure using C++ templates, which can be filled and converted to a Python Awkward Array through \verb"ak.from_buffers". The data comes out from LayoutBuilder as a set of named buffers and a JSON~\cite{json-ref} Form. The \verb"Form" is a unique description of an Awkward Array and returns a \verb"std::string" that tells Awkward Array how to put everything together. LayoutBuilder is part of an \verb"awkward-cpp" package that is separate from the \verb"awkward" package. Both packages are individually pip-installable. The code does not have helper methods to pass the data to Python, so different projects can use different binding generators. The code relies on generalized lambda expressions to deduce parameter type during compile time, which is available from the C++14 standard.

ArrayBuilder~\cite{arraybuilder-ref} and LayoutBuilder are both used to create Awkward Arrays. The main difference between a LayoutBuilder and an ArrayBuilder is that the data types that can be appended to the LayoutBuilder are defined in advance, while any data types can be appended to an ArrayBuilder. LayoutBuilder is designed to build Awkward Arrays faster. The flexibility of ArrayBuilder comes with performance limitations since it needs to discover the data type, while LayoutBuilder knows it in advance.

\section{User Interface of LayoutBuilder}

This section explains the user interface of LayoutBuilder with the help of an example of an Awkward Array with nested records and variable-length lists.

\subsection{Phases of LayoutBuilder}
There are three phases of using LayoutBuilder:
\begin{enumerate}
    \item {\bf Constructing a LayoutBuilder:} from variadic templates (It is an implicit template instantiation).
    \item {\bf Filling the LayoutBuilder:} while repeatedly walking over the raw pointers.
    \item {\bf Taking the data to user-allocated buffers:} the user can pass them to Python if needed.
\end{enumerate}

\subsection{Illustrative Example}

\begin{listing}[!ht]
\begin{minted}{cpp}
  #include "awkward/LayoutBuilder.h"
  
  enum Field : std::size_t {x, y};
  UserDefinedMap fields_map({
      {Field::x, "x"},
      {Field::y, "y"}});

  // Constructing a LayoutBuilder from variadic templates!
  RecordBuilder<
      RecordField<Field::x, NumpyBuilder<double>>,
      RecordField<Field::y, ListOffsetBuilder<int64_t, NumpyBuilder<int32_t>>>
  > builder(fields_map);
  
  auto& x_builder = builder.field<Field::x>();
  auto& y_builder = builder.field<Field::y>();

  // Filling the LayoutBuilder
  x_builder.append(1.1);
  auto& y_subbuilder = y_builder.begin_list();
  y_subbuilder.append(1);
  y_builder.end_list();
  
  x_builder.append(2.2);
  y_builder.begin_list();
  y_builder.end_list();

  x_builder.append(3.3);
  y_builder.begin_list();
  y_subbuilder.append(1);
  y_subbuilder.append(2);
  y_builder.end_list();
\end{minted}
\caption{Example of a LayoutBuilder with nested records and variable-length lists.}
\label{listing:1}
\end{listing}

An example of \verb"RecordBuilder" is illustrated in Listing~\ref{listing:1}. The first step is to include the LayoutBuilder header file (see~\cite{layoutbuilder-userguide-ref} for the installation instructions). Next, the RecordBuilder is constructed with variadic templates. The contents of a RecordBuilder are heterogeneous type containers (\verb"std::tuple") that take the other Builders as the template parameters. The field names are non-type template parameters defined by the user. Currently, it is not possible to template on strings as this functionality comes only from C++20 and onwards. Therefore, for passing the field names as template parameters to the RecordBuilder, a user-defined \verb"field_map", with enumerated type field ID as keys and the field names as value, has to be provided. In the case of multiple RecordBuilder, a user-defined map has to be specified for each of the RecordBuilder used.
 
After that, the LayoutBuilder buffers are filled with the required data as shown in Listing~\ref{listing:1}. To make sure there are no errors while filling these buffers, the user can check their validity by using the \verb"is_valid()" method, which can be called on every entry if they want to trade safety for speed. The example translates into the following Awkward Array in Python: 
\begin{minted}{python}
[{"x": 1.1, "y": [1]}, {"x": 2.2, "y": []}, {"x": 3.3, "y": [1, 2]},]
\end{minted}

We want NumPy to own the array buffers so that they get deleted when the Awkward Array goes out of Python scope, not when the LayoutBuilder goes out of C++ scope. The hand-off, therefore, needs a few steps:

\begin{enumerate}
\item Retrieve the set of buffer names and their sizes (as a number of bytes).

\begin{minted}{cpp}
  std::map<std::string, size_t> names_nbytes = {};
  builder.buffer_nbytes(names_nbytes);
\end{minted}
  
\item Allocate memory for these buffers in Python with \verb"np.empty(nbytes, type = np.uint8)" and get \verb"void*" pointers to these buffers by casting the output of \verb"numpy_array.ctypes.data".

\item Let the LayoutBuilder fill these buffers.
\begin{minted}{cpp}
  std::map<std::string, void*> buffers;
  builder.to_buffers(buffers);
\end{minted}

\item  Finally, JSON Form is generated with:
\begin{minted}{cpp}
  std::string form = builder.form();
\end{minted}
\end{enumerate} 

The Form generated for the example in Listing~\ref{listing:1} is shown in Listing~\ref{listing:2}. Now, everything can be passed over the border from C++ to Python using the \verb"py::buffer_protocol" of \verb"pybind11"~\cite{pybind11-ref} for the buffers, as well as an integer for the length and a string for the Form. If the user ever needs to make a change in the format of the records (add, remove, rename, or change the field type), there is no need to change anything in the Python-C++ interface. All of that is contained in the specialization of the C++ template and the filling procedure, which are both in the C++ code.

\begin{listing}[!ht]
\begin{verbatim}
                    {"class": "RecordArray",
                     "contents": {
                         "x": {"class": "NumpyArray",
                               "primitive": "float64",
                               "form_key": "node1"},
                         "y": {"class": "ListOffsetArray",
                               "offsets": "i64",
                               "content": {
                                   "class": "NumpyArray",
                                   "primitive": "int32",
                                   "form_key": "node3"},
                              "form_key": "node2"}" },
                     "form_key": "node0"}
\end{verbatim}
\caption{Awkward Array Form for the example in Listing~\ref{listing:1}.}
\label{listing:2}
\end{listing}

\section{Applications}

The header-only approach allows for multiple applications in both static and dynamic projects. Awkward RDataFrame~\cite{awkward-rdf-pyhep2022-ref} uses the C++ header-only libraries to simplify the process of just-in-time (JIT) compilation in \verb"ROOT"~\cite{root-ref}. The \verb"ak.from_rdataframe"~\cite{from_rdataframe-ref} function converts the selected ROOT RDataFrame~\cite{rdf-ref} columns as native Awkward Arrays. The templated header-only implementation constructs the Form from the primitive data types~\cite{awkward-to-rdf-ref}. The generation of all the types via templates makes it easier to dynamically generate LayoutBuilder from strings in Python and then compile it with \verb"cling"~\cite{cling-ref}. 

 Another application of header-only LayoutBuilder could be in the \verb"ctapipe"~\cite{ctapipe-ref} project, which is currently in the planning stage. \verb"ctapipe" is a framework for prototyping the low-level data processing algorithms for the Cherenkov Telescope Array~\cite{ctapipe-observatory-ref}. It does some processing on structured (``awkward'') event data, and the developers want to refactor their implementation to use Awkward Arrays. They already have C++ code that iterates over the custom file format, which has array types that are known at compile-time. The easiest way to Awkward Arrays in this project is to use a LayoutBuilder to fill the buffers and then send them to Python through pybind11.

\section{Conclusion}
The header-only approach presented in this paper facilitates the Python-C++ integration of Awkward Arrays and enhances their portability. A set of templated header-only libraries use only C-types (integers, strings, and raw buffers) to build Awkward Arrays and send them to Python by generating a JSON Form. A standalone awkward header-only C++ package opens up the doors for users to analyze their data in Python. Awkward Arrays can be seamlessly integrated with external projects without linking against platform-specific libraries or worrying about native dependencies. This new development also allows the extension of the use cases of Awkward Arrays to scientific communities beyond HEP.

\section{Acknowledgment}
This work is supported by the NSF cooperative agreement OAC-1836650 (IRIS-HEP) and the NSF cooperative agreement PHY-2121686 (US-CMS LHC Ops).


\section*{References}
\begin{thebibliography}{9}
\bibitem{awkward-ref}
Pivarski J, Osborne I, Ifrim I, Schreiner H, Hollands A, Biswas A, Das P, Roy Choudhury S, Smith N and Goyal M 2018 Awkward Array [Computer software] {\it Zenodo}
\verb"https://doi.org/10.5281/zenodo.4341376"

\bibitem{sci-kit-hep-ref} Rodrigues E 2019 The Scikit-HEP Project {\it EPJ Web Conf.} {\bf 214} 06005 
\verb"DOI:10.1051/epjconf/201921406005"

\bibitem{numpy-ref} Harris C R, Millman K J, van der Walt S J et al. 2020 Array programming with NumPy {\it Nature} {\bf 585} 357–362. \verb"DOI: 10.1038/s41586-020-2649-2"

 \bibitem{columnar-ref} Pivarski J, Nandi J, Lange D and Elmer P, 2019 Columnar data processing for HEP analysis {\it EPJ Web Conf.} {\bf 214} 06026
 \verb"DOI: 10.1051/epjconf/201921406026"

 \bibitem{awkward-paper-ref} Pivarski J, Elmer P and Lange D 2020 Awkward Arrays in Python, C++, and Numba {\it EPJ Web Conf.} {\bf 245} 05023
\verb"DOI: 10.1051/epjconf/202024505023"

\bibitem{awkward-scipy2020-ref} Pivarski J, Osborne I, Das P, Biswas A and Elmer P 2020 Awkward Array: JSON-like data, NumPy-like idioms {\it Proc. of the 19th Python in Science Conf. (SCIPY 2020)} 78-84

\bibitem{awkward-lessonslearnt-ref} Pivarski J 2021 Lessons learned in Python-C++ integration {\it 20th International Workshop on Advanced Computing and Analysis Techniques in Physics Research} \verb"https://indi.to/N69ds"

\bibitem{data-science-tools-ref} Pivarski J, Rodrigues E, Pedro K, Shadura O, Krikler B and Stewart G A 2022 HL-LHC Computing Review Stage 2, Common Software Projects: Data Science Tools for Analysis {\it [arXiv 2202.02194]}
\verb"DOI: 10.48550/arXiv.2202.02194"

\bibitem{layoutbuilder-userguide-ref} User Guide: How to Use Header-Only LayoutBuilder in C++ {\it Awkward Array Documentation}
\verb"https://awkward-array.org/doc/main/user-guide/how-to-use-header-only-layoutbuilder.html"

\bibitem{json-ref} JSON \verb"https://www.json.org/"

\bibitem{arraybuilder-ref} ArrayBuilder \verb"https://awkward-array.org/doc/main/reference/generated/ak.ArrayBuilder.html"

\bibitem{pybind11-ref} Jakob W, Rhinelander J and Moldovan D 2016 pybind11 - Seamless operability between C++11 and Python.
\verb"https://github.com/pybind/pybind11"

\bibitem{awkward-rdf-pyhep2022-ref} Osborne I and Pivarski J 2022 Awkward RDataFrame Tutorial {\it PyHEP 2022 (virtual) Workshop}
\verb"https://doi.org/10.5281/zenodo.7081586"

\bibitem{root-ref} Brun R, Rademakers F, Canal P, Naumann A, Couet O, Moneta L, Vassilev V, Linev S, Piparo D, GANIS G, Bellenot B, Guiraud E, Amadio G, wverkerke, Mato P, TimurP, Tadel M, wlav, Tejedor E, Blomer J, Gheata A, Hageboeck S, Roiser S, marsupial, Wunsch S, Shadura O, Bose A, CristinaCristescu, Valls X and Isemann R 2019 root-project/root: v6.18/02 (v6-18-02) {\it Zenodo} \verb"https://doi.org/10.5281/zenodo.3895860"

\bibitem{from_rdataframe-ref} from\_rdataframe \verb"https://awkward-array.org/doc/main/reference/generated/ak.from_rdataframe.html"

\bibitem{rdf-ref} Piparo D, Canal P, Guiraud E, Pla X V, Ganis G, Amadio G, Naumann A and Tejedor E 2018 RDataFrame: Easy Parallel ROOT Analysis at 100 Threads {\it EPJ Web Conf.} {\bf 214} 06029
\verb"DOI: 10.1051/epjconf/201921406029"

\bibitem{awkward-to-rdf-ref} Osborne I and Pivarski J 2022 Awkward to RDataFrame and back {\it [arXiv 2302.09860]}
\verb"DOI: 10.48550/arXiv.2302.09860"

\bibitem{cling-ref} Cling [Online]. \verb"https://root.cern.ch/cling"

\bibitem{ctapipe-ref} ctapipe documentation \verb"https://ctapipe.readthedocs.io/en/latest/"

\bibitem{ctapipe-observatory-ref} ctapipe observatory \verb"https://www.cta-observatory.org/"

\end{thebibliography}
\end{document}